\documentclass[prd,twocolumn,showpacs,reprint,preprintnumbers,nofootinbib]{revtex4-1}

\usepackage[font=small,labelfont=bf,justification=centerlast]{caption}
\usepackage{array,multirow,makecell}
\usepackage{subcaption}
\usepackage{slashed}
\usepackage{hypernat}

\usepackage{graphicx}
\usepackage[nointegrals]{wasysym}
\usepackage{rotating}
\usepackage{color}
\usepackage{epsfig}
\usepackage{bm}
\usepackage{hyperref}
\usepackage{url}
\usepackage{slashed}
\usepackage{amsmath}
\usepackage{amssymb}
\usepackage[dvipsnames]{xcolor}

\usepackage{geometry}
\newcommand{\GeV}{{\rm GeV}}

\newcommand{\cm}{{\rm cm}}

\newcommand{\be}{\begin{equation}}
\newcommand{\ee}{\end{equation}}
\newcommand{\bea}{\begin{eqnarray}}
\newcommand{\eea}{\end{eqnarray}}
\newcommand{\la}{\left\langle}
\newcommand{\ra}{\right\rangle}

\newcommand{\MM}{\left|\mathcal{M}\right|^2}

\def\lg{{\mathchoice{~\raise.58ex\hbox{$<$}\mkern-14.8mu\lower.52ex\hbox{$>$}~}
                    {~\raise.58ex\hbox{$<$}\mkern-14.8mu\lower.52ex\hbox{$>$}~}
                    {\raise.59ex\hbox{{$\scriptscriptstyle <$}}\mkern-12.8mu%
                     \lower.01ex\hbox{{$\scriptscriptstyle >$}}}   {}   }}
\def\gl{{\mathchoice{~\raise.58ex\hbox{$>$}\mkern-12.8mu\lower.52ex\hbox{$<$}~}
                    {~\raise.58ex\hbox{$>$}\mkern-12.8mu\lower.52ex\hbox{$<$}~}
                    {\raise.62ex\hbox{{$\scriptscriptstyle >$}}\mkern-12.0mu%
                     \lower.05ex\hbox{{$\scriptscriptstyle <$}}}  {}    }}

\begin{document}

\title{Impact of dark matter self-scattering on its relic abundance
}

\author{Andrzej Hryczuk}
\email{andrzej.hryczuk@ncbj.gov.pl}
\author{Maxim Laletin}
\email{maxim.laletin@ncbj.gov.pl}
\affiliation{National Centre for Nuclear Research, Pasteura 7, 02-093 Warsaw, Poland}

\date{\today}

\begin{abstract}
Elastic self-scatterings do not change the number of dark matter particles and as such have been neglected in the calculation of its relic abundance. In this work we highlight the scenarios where the presence of self-scatterings has a significant impact
on the effectiveness of annihilation processes through the modification of dark matter momentum distribution. We study a few example freeze-out scenarios involving resonant and sub-threshold annihilations, as well as a model with an additional source of dark matter particles from the decays of a heavier mediator state. Interestingly, when the calculation is performed at the level of dark matter momentum distribution function, we find that the injection of additional energetic dark matter particles onto the thermal population can lead to a \textit{decrease} of its final relic abundance. 
\end{abstract}

\maketitle

\section{Introduction}

The extremely dense and hot plasma of the very Early Universe exhibits exceptional conditions for particle creation. Every new particle species that \textit{i)} is coupled to the Standard Model (SM) states,  \textit{ii)} has a non-vanishing mass not exceeding the reheating temperature and \textit{iii)} is stable on cosmological timescales, will inevitably end up with a relic thermal population contributing to the present-day dark matter (DM) density. In the simplest scenarios this thermal component accounts for all of the observed DM, with $\Omega h^2 = 0.120 \pm 0.001$ \cite{Planck:2018vyg}. If adopted, this requirement puts significant constraints on the parameters of a given DM model that affect the rates of the particle-number-changing processes.

An attractive, yet not overly restrictive assumption is that the interactions between the DM particles and SM plasma are sufficiently frequent to enforce chemical equilibrium at some time in the very Early Universe. In such scenario the relic density of DM is predominantly determined by just one quantity – the cross section of annihilation – in a wide range of DM masses. Effectively, when the rate of annihilation drops below the rate of the Universe expansion, the dark sector departs from the chemical equilibrium with the SM plasma and the number of DM particles in the comoving volume ceases to change with time (\textit{freezes-out}), hence establishing the relic population. The general approach to determine the relic abundance of DM is solving the Boltzmann equation (BE) that describes the evolution of the DM distribution function in the expanding Universe. To calculate the relic abundance often only the 0th moment of this equation, tracing only the particle-number density, is considered, which is what is used in numerous existing numerical packages, e.g.  \cite{Belanger:2018ccd,Bringmann:2018lay,Ambrogi:2018jqj}.

In a broader class of DM models thermodynamics in the early Universe can be more elaborate and the relic abundance can be generated in other ways than described above, but the leading principle that the relic density is determined by the interplay of DM number-changing processes remains. 

However, the rate of number-changing processes depends not only on the interaction strength, but also on the characteristics of the DM population, in particular its number density and momentum distribution. While the former is determined by chemical equilibration and decoupling governed by annihilations, the latter is related to the local kinetic equilibrium that is maintained mostly by elastic and inelastic scatterings on the particles from the SM plasma.
Although in the typical models of weakly interacting massive particles (WIMPs) the kinetic equilibrium is maintained long after the freeze out, exceptions to this standard scenario exist in even simple models \cite{vandenAarssen:2012ag,Kuflik:2015isi,Binder:2017rgn,Garny:2017rxs,Hektor:2019ote,Abe:2020obo,Abe:2021jcz,Binder:2021bmg} and are expected to occur much more often in more involved scenarios containing processes actively disrupting local thermal equilibrium, e.g. decays of heavier states or self-heating \cite{Kamada:2017gfc} due to semi-annihilations, cannibalization \cite{Carlson:1992fn} or conversions (see e.g. \cite{Garny:2017rxs,Ivanov:2018srm,Ghosh:2021qbo}).
In such cases the crucial question is whether the scattering processes are efficient enough to force the momentum distribution to follow the equilibrium one with the same temperature as the SM plasma or not.

The impact of elastic scatterings between DM and the thermal-bath particles has been studied in recent years both at the level of tracing the evolution of DM temperature, see e.g. \cite{Arcadi:2011ev,vandenAarssen:2012ag,Kuflik:2015isi,Binder:2017rgn,Duch:2017nbe,Yang:2019bvg,Fitzpatrick:2020vba}, alongside the number density in a coupled systems of Boltzmann equations (cBE) for the 0th and the 2nd moments of the distribution function, as well as at the level of numerical solution of the full momentum-dependent Boltzmann equation (fBE) \cite{Binder:2017rgn,Binder:2021bmg,Du:2021jcj,Ala-Mattinen:2022nuj}. When comparing these two approaches it was noted \cite{Binder:2017rgn} that although cBE and fBE give in general consistent results, they can substantially differ in scenarios with strong velocity-dependent annihilations, e.g. due to a resonance or a threshold. This is because cBE enforces the shape of the distribution to be of the Maxwell-Boltzmann form, which is equivalent to insisting that DM self-scatterings, redistributing energy in the dark matter component, are very efficient. While without the explicit inclusion of DM self-scattering on top of annihilation and elastic scatterings on the thermal-bath particles, the fBE method effectively neglects all such processes altogether. 
Hence, which of these approaches gives a better estimate clearly depends on the actual strength of self-scattering processes.

In this paper we implement, for the first time in the literature, the complete DM self-scatterings at the level of fBE in the context of thermal relic density calculation
\footnote{For an implementation within a relaxation-like approximation see Ref.~\cite{Ala-Mattinen:2019mpa}.}
and investigate how do they modify the energy distribution of DM and ultimately its relic density for three different models of DM with strong velocity dependence of annihilation processes. We compare the results of the different approaches discussed above to shed light on their ranges of applicability regarding the rate of self-scattering. In particular, we explore a model with the injection of an energetic component of DM in the form of heavier particle decay products. 

It is worth mentioning, that as far as elastic scatterings on thermal-bath particles and annihilations are typically strongly tied together in the underlying DM model, the self-scattering processes are often unrelated and in particular can have much larger cross sections. In fact, models of strongly interacting DM have gained a substantial attention in the literature as a possible solution of the discrepancies between observations and theoretical prediction for the density profiles of DM in small scale structures (see e.g. Ref.~\cite{Tulin:2017ara} for a review).

The paper is organized as follows. In Sec.~\ref{sec:self} we describe our implementation of self-scattering processes in the Boltzmann equation. Sec.~\ref{sec:ImpactFO} discusses applications to two models with the standard freeze-out production, while in Sec.~\ref{sec:decay} we extend the analysis to an example with an additional non-thermally produced component and its interplay with the freeze-out mechanism. Finally, Sec.~\ref{sec:conclusions} concludes.

\section{Dark matter self-scattering}
\label{sec:self}

Before we discuss the specifics of self-scattering processes, 
let us briefly review the formalism used in the calculations of the thermal relic density. The evolution of the DM component is well described in the semi-classical limit by the Boltzmann equation that in Friedmann-Roberston-Walker space-time takes the following form:
\begin{eqnarray}
  \label{diff_boltzmann}
  E\left(\partial_t-Hp\partial_p\right)f_\chi &=& C_{\rm ann}[f_\chi] \; + \; C_{\rm el}[f_\chi] \;  \nonumber \\
  &+& C_{\rm dec}[f_\chi] \; + \; C_{\rm self}[f_\chi]\,,
\end{eqnarray}
where $f_\chi(t,p)$ is the DM distribution function depending on time $t$ (in what follows replaced with temperature $T$) and momentum $p$ (or equivalently, energy $E$), $H$ stands for the Hubble expansion rate and $C$ denotes different collision terms that are relevant in the Early Universe. 
As an example we provide here the structure of the collision term for a particle $i$ that participates in a general 2-to-2 process
\begin{multline}
  \label{C2to2_def}
  C_{ij \rightarrow mn} =\frac{1}{2g_i}\int\!\!\frac{d^3\tilde p}{(2\pi)^32\tilde E}\int\!\!\frac{d^3k}{(2\pi)^32\omega}\int\!\!\frac{d^3\tilde k}{(2\pi)^32\tilde \omega} \\
  \times (2\pi)^4\delta^{(4)}(\tilde p+p-\tilde k-k)\\
\times\Big[
\left|\mathcal{M}\right|^2_{ij \leftarrow mn}f_m(\omega)f_n(\tilde \omega)[1 \pm f_i(E)][1 \pm f_j(\tilde E)] \\
-\left|\mathcal{M}\right|^2_{ij \rightarrow mn}f_i(E)f_j(\tilde E)[1 \pm f_m(\omega)][1 \pm f_n(\tilde \omega)]
\Big],
\end{multline}
where $g_i$ is the number of degrees of freedom of particle $i$, 
$\omega$ and $\tilde{\omega}$ (4-momenta $k$ and $\tilde{k}$) are the energies of the final state particles $m$ and $n$, and $E$ and $\tilde{E}$ are the energies of the initial state particles $i$ and $j$ respectively (4-momenta $p$ and $\tilde{p}$). The amplitude squared $\MM$ is summed over both initial and final internal degrees of freedom. The corresponding distribution functions are marked with the respective indices. The first term in this expression is generally referred to as the gain term, while the second – the loss term. The signs in front of the distribution functions of the products in the forward and backward reactions depend on the spin statistics describing these particles. If these particles constitute a very dilute gas $[1 \pm f] \approx 1$ and these factors can be neglected. 

The standard approach \cite{Gondolo:1990dk} of solving only the single Boltzmann equation for the number density $n_\chi$ (nBE) can be obtained from Eq.~\eqref{diff_boltzmann} by the integration over the momentum $\vec{p}$ leading to 
\be
\frac{dn_\chi}{dt} + 3Hn_\chi = g_\chi \int\! \frac{d^3p}{(2\pi)^3 \, E} \{ \, C_{\rm ann}[f_\chi] +  C_{\rm dec}[f_\chi] \, \}.
\ee
The collision terms for elastic and self-scattering do not change the number density and therefore cancel out after the integration. However, in order to solve this equation one needs to know the form of $f_\chi(t,p)$, so that the r.h.s. of the equation can be integrated and expressed in terms of the number density. A common assumption is that the distribution of DM has an equilibrium shape (Fermi-Dirac/Bose-Einstein distributions in general or Maxwell-Boltzmann distribution in the dilute or non-relativistic limit) that corresponds to the temperature of the SM plasma and with a potentially non-zero chemical potential that is effectively solved for. This assumption is often justified since the elastic scattering processes on SM particles typically proceed at a large enough rate. In cases when the elastic scatterings cannot maintain local thermal equilibrium, but the shape of the distribution function is still close to the thermal one, albeit with $T_\chi\neq T$, the cBE approach is expected to give an accurate prediction not only for the DM relic abundance, but also for its temperature evolution. This system of equations for the number density and temperature is obtained from Eq.~\eqref{diff_boltzmann} by the integration over $(g_\chi/(2\pi)^3) \int d^3p/E $ and $(g_\chi/(2\pi)^3) \int d^3p \; p^2/E^2 $ respectively. In the following we show the results obtained with these two approaches only as a comparison to the full treatment we study in this work. Thus, for more technical details regarding nBE and cBE we refer to Ref.~\cite{Binder:2017rgn}, while below we discuss our implementation of the fBE, and especially the self-scatterings.

The collision term for annihilation of DM particles into two SM states (neglecting the $\left[1 \pm f_\chi\right]$ factors) is given by
\begin{multline}
\label{Candd_def}
C_\mathrm{ann}=\frac{1}{2g_\chi}\int\frac{d^3\tilde p}{(2\pi)^32\tilde E}\int\frac{d^3k}{(2\pi)^32\omega}\int\frac{d^3\tilde k}{(2\pi)^32\tilde \omega}\\
\times(2\pi)^4 \, \delta^{(4)}(\tilde p+p-\tilde k-k)\\
\times\big[\left|\mathcal{M}\right|^2_{\bar\chi\chi\leftarrow \bar f f}g(\omega)g(\tilde \omega) -\left|\mathcal{M}\right|^2_{\bar\chi\chi\rightarrow \bar f f}f_\chi(E)f_\chi(\tilde E)
\big]
\end{multline}
where $g$ stands for the thermal distribution for a considered thermal-bath state. The gain term in the annihilation collision term 
does not contain any unknown distribution functions and in principle can be calculated explicitly. 
The loss term contains two DM distribution functions: one of them can be taken out of the integration, the other has to be integrated over the corresponding momentum, while the residual part of the integrand can be expressed in terms of annihilation cross section. To perform this integration numerically the unknown distribution function can be regarded as a combination of discrete components $f_{\chi}(p_i)$ for a given value of momentum $p_i$. Thus, the momentum-dependent Boltzmann equation is split into a system of ordinary differential equations for each momentum component and the integration is approximated with a weighted sum of these components. This scheme in particular is realized in the DRAKE code \cite{Binder:2021bmg}, which we use for the solution of the BE, except for the self-scatterings (see below) which are not included in the current public version of DRAKE. 

The same procedure can be in principle applied to the loss term of the elastic collision term
\begin{multline}
\label{Celd_ef}
C_\mathrm{el}=\frac{1}{2g_\chi}\int\frac{d^3\tilde p}{(2\pi)^32\tilde E}\int\frac{d^3k}{(2\pi)^32\omega}\int\frac{d^3\tilde k}{(2\pi)^32\tilde \omega}\\
\times(2\pi)^4 \, \delta^{(4)}(\tilde p+\tilde k-p-k) \, {\left|\mathcal{M}\right|}^2_{\chi f\leftrightarrow\chi f}\\
\times\big[\left(1\mp g^\pm(\omega)\right)\, g^\pm(\tilde\omega)f_\chi({\tilde E})- (\omega\leftrightarrow\tilde\omega, {E}\leftrightarrow{\tilde E})\big]\,,
\end{multline}
which has the same structure as the gain term, but with the energies transformed as indicated. 
However, in the limit of small momentum transfer the whole $C_{\rm el}$ can be expressed through the DM distribution function and its derivatives without any numerical integrations left~\cite{Bringmann:2006mu} (see Appendix~\ref{append_B}). Finally, the decay collision term $C_{\rm dec}$ can be simplified to an analytical expression (see Sec.~\ref{sec:decay}), as long as the decaying particle is described by an equilibrium distribution.

The collision term for self-scattering has the following general expression (neglecting the $\left[1 \pm f_\chi\right]$ factors)
\begin{widetext}
\bea
  \label{Cself_def}
  C_\mathrm{self}&=&\frac{1}{2g_\chi}\int\frac{d^3\tilde p}{(2\pi)^32\tilde E}\int\frac{d^3k}{(2\pi)^32\omega}\int\frac{d^3\tilde k}{(2\pi)^32\tilde \omega} \quad (2\pi)^4\delta^{(4)}(\tilde p+p-\tilde k-k)\nonumber\\
&&\times \Big\{ \frac{1}{2} \left|\mathcal{M}\right|^2_{\chi\chi \leftrightarrow \chi\chi} \left[
f_\chi(\omega)f_\chi(\tilde \omega) - 
f_\chi(E)f_\chi(\tilde E) 
\right] + \left|\mathcal{M}\right|^2_{\chi\bar{\chi} \leftrightarrow \chi\bar{\chi}} \left[
f_\chi(\omega)f_{\bar{\chi}}(\tilde \omega) - 
f_\chi(E)f_{\bar{\chi}}(\tilde E) 
\right] \Big\}.
\eea
\end{widetext}
The first term (both loss and gain parts) describes the scattering on particles and the second term the scattering on antiparticles. The factor $1/2$ in front of the first one takes into account the symmetry between the identical particles $\chi$ with momenta that are integrated over. In the absence of $CP$-violating processes in the dark sector the distribution function for particles and antiparticles is always the same, thus the self-scattering collision term can be written in terms of an effective amplitude squared 
\be
\label{M2_self}
\MM_{\rm self} = \frac{1}{2} \left|\mathcal{M}\right|^2_{\chi\chi \leftrightarrow \chi\chi} + \left|\mathcal{M}\right|^2_{\chi\bar{\chi} \leftrightarrow \chi\bar{\chi}} ,
\ee
and one set of gain and loss terms.  

In comparison to elastic scattering, the rate of self-scattering is suppressed by an additional $f_\chi$ in the collision term, especially with respect to light SM states that are greatly more abundant in equilibrium. However, it is an insufficient reason to claim that self-scatterings play a little role in shaping the DM energy distribution. First of all, an average relative momentum transfer for elastic scattering is $\delta p/p \sim (T/m_{\chi})^{1/2} \ll 1$, while for self-scattering $\delta p/p \sim 1$, so an effective energy redistribution in the latter case does not require many collisions. It is useful to compare the characteristic relaxation times for both processes $\tau_r \sim N_{\rm coll}/\Gamma$, where $\Gamma = n_\chi \la \sigma v\ra$ is the rate of scatterings 
and $N_{\rm coll}$ is the number of collisions required to substantially change the momentum of DM (see also Ref.~\cite{Bringmann:2006mu}). In case of elastic scatterings $N^{\rm el}_{\rm coll} \sim m_{\chi}/T$, the density of relativistic SM particles $n_{\rm SM} \propto T^3$ and the overall relaxation time scales with temperature as $T^{-6}$ \cite{Hofmann:2001bi}. For self-scatterings after freeze-out the density of DM particles scales with temperature by the same law due to the expansion, but $N^{\rm self}_{\rm coll} \sim 1$ and for the models that we consider in Sec.~\ref{sec:VRES} and \ref{sec:decay} $\la \sigma v\ra_{\rm self} \propto T$ in the non-relativistic limit. Thus, the relaxation time for self-scattering scales with temperature as $T^{-4}$, which means that self-scattering processes remain an effective mean of equilibration longer than elastic scatterings. 

Secondly, self-scattering can rely on different couplings (or even on a different type of interaction) than the ones that govern elastic scattering. For instance, in the case of scalar DM model considered in Sec.~\ref{sec:TH} self-interaction can arise from a simple $\phi^4$ vertex interaction with the amplitude squared proportional to the square of the respective coupling, while elastic scattering on SM fermions is loop-suppressed. In the case of fermion DM coupled to a vector mediator (Sec.~\ref{sec:VRES}), the rate of elastic scattering can be suppressed by the squared ratio of the two couplings, given that the coupling of the mediator to the heat bath fermions is smaller. In other models of DM self-scattering can be boosted w.r.t. elastic scattering by an $s$-channel resonance, Sommerfeld enhancement, etc. In addition, DM self-interactions are generally not as constrained by observations as the elastic scatterings. For example, the upper bound on the cross section of electron scattering for a DM particle with the mass of 1 GeV is $\sim 10^{-34} \cm^2$ \cite{Nguyen_2021}, while the cross section of DM self-scattering for that mass can be as large as $\sim 10^{-24} \cm^2$ \cite{10.1093/mnras/stx896}.

From a technical point of view, the additional complication introduced by the self-scattering collision term in Eq.~\ref{Cself_def} comes from the gain term that contains two distribution functions of the products. 
While the loss term can be treated in the same way as for annihilation, the gain term for self-scattering cannot be simply formulated in terms of the cross section and the presence of two unknown function in the integrand leads to complicated angular dependencies during the integration (see Appendix~\ref{append_A}). Since self-scatterings are not implemented in the current version of DRAKE, we merged the existing version with the program for the calculation of the self-scattering collision term that we developed. 
Though the loss term in Eq.~\ref{Cself_def} can be expressed through the self-scattering cross section and implemented numerically with one summation of the momentum components, we use the same procedure as for the gain term to achieve a better numerical cancellation between the two terms close to equilibrium point, since the same interpolation procedure is used in both cases. 
 
\section{Impact in freeze-out models}
\label{sec:ImpactFO}

During the freeze-out process the self-scattering does not introduce any direct change in the DM number density, but indirectly it can significantly affect the annihilation rates. To exemplify this we have chosen to present results of a study of two models introduced in Ref.~\cite{Binder:2021bmg}: the generic vector resonance and sub-threshold scenarios. In both cases the kinetic decoupling and non-equilibrium shape of $f_\chi(p)$ can have a strong impact on the final relic abundance and as noted in Ref.~\cite{Binder:2021bmg} there can be a substantial difference between the cBE and fBE approaches. As it was pointed out in Ref.~\cite{Binder:2017rgn} it is expected that inclusion of self-scatterings to fBE treatment should in the limit of large self-interactions lead to result coinciding with cBE. In this section we demonstrate that this is indeed the case and quantify how strong the self interactions need to be to have an impact. 

\subsection{Vector resonance model}
\label{sec:VRES}

The arguably most common scenario where the DM annihilation cross-section has a strong velocity dependence arises in models with $s$-channel resonance. For concreteness let us take exactly the same model as in Ref.~\cite{Binder:2021bmg} where the resonance is mediated through an exchange of 
a generic vector mediator $A^{\mu}$, with the interaction Lagrangian
\begin{align}
\mathcal{L} \supset - \lambda_{\chi} \bar{\chi} \gamma^{\mu} \chi A_{\mu} - \lambda_f \bar{f} \gamma^{\mu} f A_{\mu}\,.
\label{eq:Lres}
\end{align}
The model can be described by a set of five parameters: the DM mass $m_{\chi}$, the mediator mass $m_A$, the mass ratio of heat-bath fermions to DM $r\equiv m_f/m_{\chi}$ and finally the coupling constants $\lambda_f$ and $\lambda_{\chi}$. Out of these input parameters it is convenient to define deviation from the exact resonance position $\delta \equiv (2m_{\chi}/m_A)^2-1$ and a dimensionless measure of the total decay width of $A^\mu$, $\tilde\gamma \equiv \Gamma_A/m_A$. Note, that compared to the discussion in Ref.~\cite{Binder:2021bmg} we separate the couplings $\lambda_f$ and $\lambda_{\chi}$, as the self-interactions break the degeneracy between them in the calculation of the DM annihilation cross section.  

Indeed, the annihilation cross-section for the process $\chi \bar{\chi} \rightarrow A^{\star} \rightarrow f \bar{f}$, can be written as~\cite{Binder:2021bmg}
\begin{equation}
\sigma v_{\rm lab} =\frac{\lambda_\chi^2 \lambda_f^2 }{384\pi m_\chi^2}\frac{(1-r^2/\tilde s)^{1/2}(1+\delta)^2 }{2\tilde s -1 }  \alpha(\tilde s)D(\tilde s)\,,
\end{equation}
with
$\alpha(\tilde s)=4(2\tilde s +1)(2\tilde s +r^2)$ and $\tilde{s} \equiv s/(4m_\chi^2)$,
where $\sqrt{s}$ is the center-of-mass energy, and
\begin{equation}
\label{eq:Ds}
D(\tilde s) \equiv \frac{1}{\left[\tilde s (1+\delta)-1\right]^2+\tilde\gamma^2}
\end{equation}
being the Breit-Wigner propagator.\footnote{Note however, that the Breit-Wigner form might not be sufficient in some cases, in which the velocity-dependent width should be used instead \cite{Duch:2017nbe}.} 
The self-scattering amplitude squared in this model consists of two contributions, as in Eq.~\eqref{M2_self}, but both of them depend only on the coupling $\lambda_\chi$
\begin{multline}
\label{M2self_res}
\MM_{\rm self} = \lambda_{\chi}^4 \, (\delta +1)^2 \times \\ 
\times \Big[ 2\beta_1(\tilde{s},\tilde{t}) \,  D(\tilde{s})D(\tilde{t}) + \beta_2(\tilde{s},\tilde{t}) D(\tilde{s})D(\tilde{u}) \Big] , 
\end{multline}
where $\tilde{t} \equiv t/(4m_\chi^2)$, $\tilde{u} = 1 - \tilde{s} - \tilde{t}$ and $\beta_1$ and $\beta_2$ are functions of $\tilde{s}$ and $\tilde{t}$ defined in the Appendix~\ref{append_B}.

In the calculations we also include elastic scatterings on the thermal-plasma fermions $f$,
which is exactly in the same way as in Ref.~\cite{Binder:2021bmg}, to where we refer the reader for more details.

\begin{figure}
    \centering
    \includegraphics[scale=0.673]{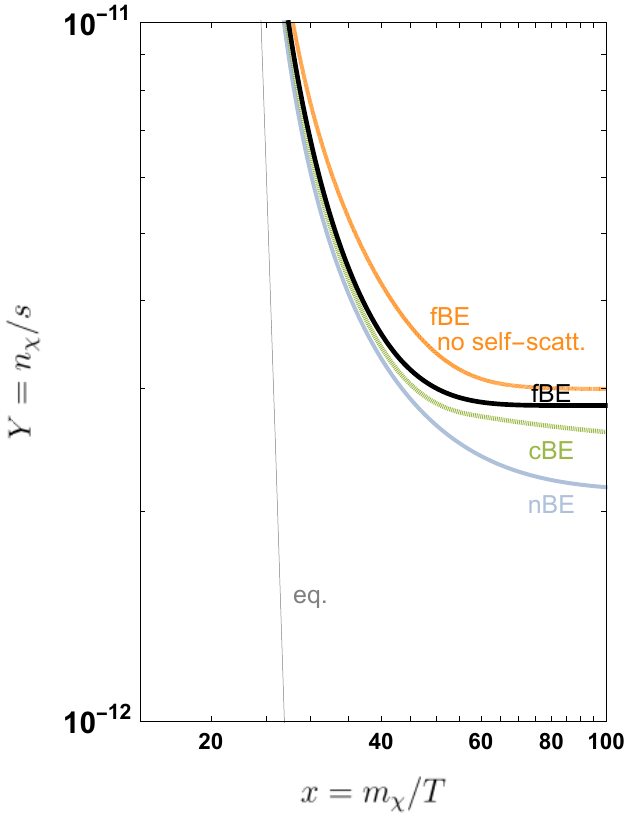}
    \hspace{-0.2cm}
    \includegraphics[scale=0.61]{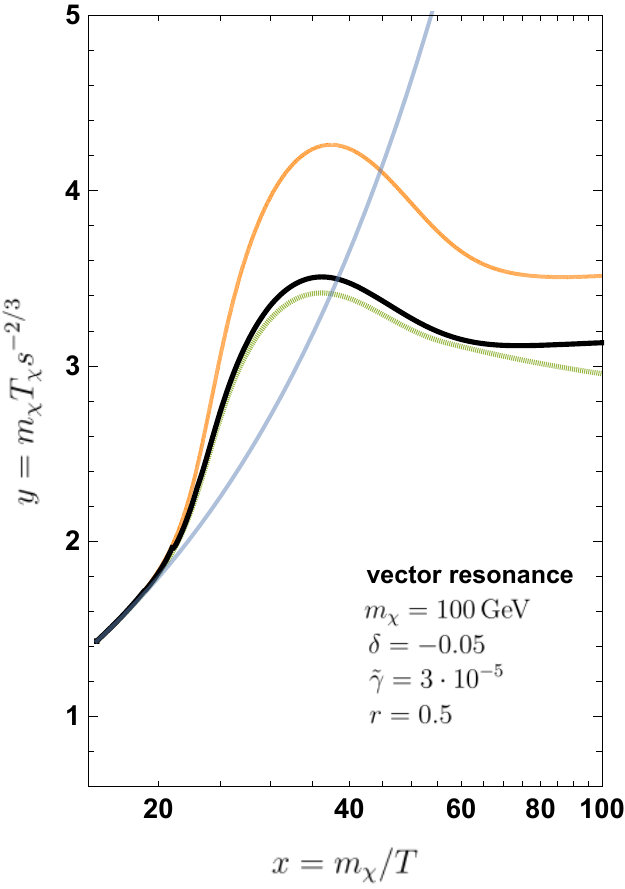}
    \caption{An example evolution of the particle yield $Y$ (left panel) and the temperature parameter $y$ (right panel) for the benchmark vector resonance model. On both panels the blue line gives the result with nBE treatment, green the cBE, orange fBE without self-scatterings, and black of the full calculation. The couplings are fixed by the value of $\tilde \gamma$ and the requirement that $(\Omega h^2)_{\rm nBE}=0.12$.}
    \label{fig:VRES_Yy}
\end{figure}

\begin{figure}
    \centering
    \includegraphics[scale=0.74]{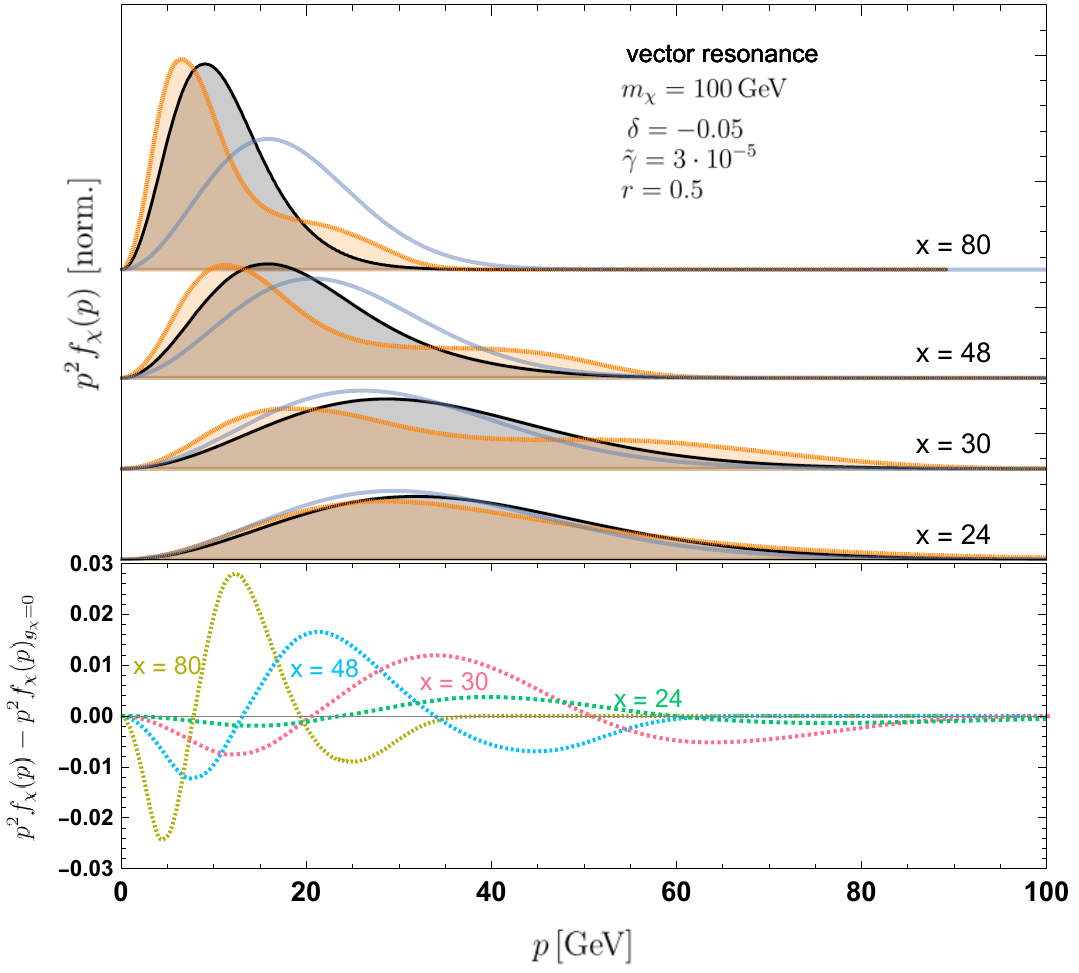}
    \caption{\textit{Top panel}: time snapshots around the freeze-out of the evolution of the normalized momentum distribution for the benchmark vector resonance model. Black (orange) points show $f_\chi(p)$ with (without) self-scatterings, while for comparison the blue line shows the equilibrium distribution at the SM plasma temperature. \textit{Bottom panel}: the difference between the normalized $f_\chi(p)$ with and without the inclusion of self-scatterings, to highlight the size of the deviation.}
    \label{fig:VRES_f}
\end{figure}

Let us start the discussion of the results from presenting the evolution of the yield $Y=n_\chi/s$ and temperature parameter 
 $y=m_{\chi} T_\chi s^{-2/3}$, where $s(T)$ is the entropy density, for a benchmark point with the resonance having SM Higgs-like width $\tilde \gamma = 3\times 10^{-5}$, and relatively heavy annihilation products, $r=0.5$ with $m_{\chi}=100$ GeV. Fixing the width $\tilde \gamma$ and requiring that the nBE solution provides the observed relic abundance defines the couplings $\lambda_f=10^{-3}$
and $\lambda_{\chi}=5.85\times 10^{-2}$
and ultimately the strength of self-scattering process. The result of the evolution is given in Fig.~\ref{fig:VRES_Yy}  for the nBE (blue), cBE (green), fBE without (orange) and including self-scatterings (black). For this benchmark $\delta=-0.05$ and thus the resonant annihilations deplete momenta around the peak of the distribution at the time of freeze-out, resulting in a slight temperature raise at first and then an abrupt chemical and kinetic decoupling. Self-scatterings reshuffle the DM particles' momenta, re-populate the regions depleted by annihilation and thus prolong the freeze-out, leading to a lower final abundance. This can be directly seen in Fig.~\ref{fig:VRES_f}, where the four time snapshots of the momentum distribution are given. Comparing to the result obtained with self-scatterings (black), which make $f_\chi(p)$ retain shape close to thermal, the curves without self-scattering (orange) show a significant dip in the distribution for the momenta that are slightly above the peak of the distribution. The particles with these momenta are efficiently depleted by the resonant annihilation, while elastic scatterings on the thermal bath are not sufficient to replenish them effectively.
The overall distribution is also visibly shifted from the equilibrium one (blue) indicating that $T_{\chi}$ is significantly lower than $T$ indeed. The bottom panel of Fig.~\ref{fig:VRES_f} highlights the relative size of the difference between the two distributions, with different colours signifying different time snapshots.

\begin{figure}
    \centering
    \includegraphics[scale=0.9]{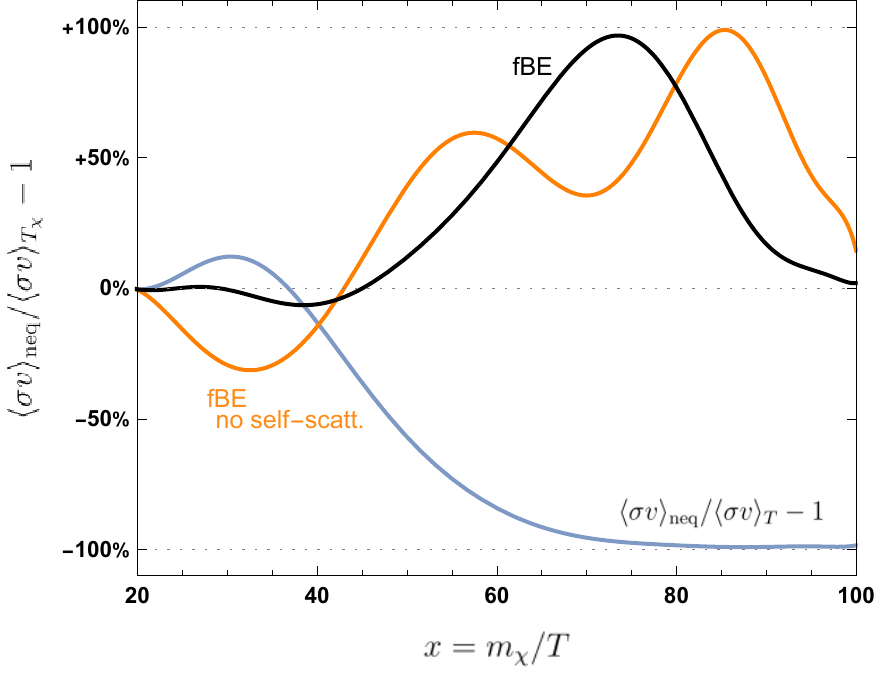}
    \caption{
    Effect of the deviation from the thermal shape of $f_\chi(p)$ on the thermally-averaged annihilation cross section for the same benchmark vector resonance model and $\lambda_\chi = 0.0585$. To demonstrate the impact of only the shape modification the black (orange) line shows the difference of $\langle \sigma v\rangle_{\rm neq}$ with (without) self-scatterings compared to $\langle \sigma v\rangle_{T_{\chi}}$ calculated at $T_{\chi}$ corresponding to the given fBE solution. This can be contrasted with the blue line that includes the impact of the temperature change as well, where the comparison is made to $\langle \sigma v\rangle_{T}$.
 }
    \label{fig:VRES_sv}
\end{figure}

One can see that the self-thermalization due to self-scatterings in this model is rather efficient, in large part due to the fact that the distortion introduced by $v$-dependent annihilation is limited. Nevertheless, even such relatively small deviation from thermal shape visibly affects the relic abundance through modifying the annihilation rate. Quantitatively, Fig.~\ref{fig:VRES_sv} shows the relative difference between the thermally averaged cross sections $\langle \sigma v\rangle$ calculated with the resulting $f_\chi(p)$ and with a thermal distribution of the same temperature. The result without self-scatterings (orange line) around the freeze-out time $x\sim 25$ shows significant deviation from the thermal one, while with self-scatterings (black line) a much milder one. When the temperature drops, to about $x\sim 50$ and further, both solutions predict an enhanced annihilation rate, but with somewhat different dependence. In particular the peak of this enhancement happens earlier for the solution with self-scatterings than without, which is a consequence of an impact on resonant annihilation by an interplay between the temperature drop and the distribution shape modification. The deviations from unity thus show that even with the self-scatterings the annihilations introduce too much of a disruption to allow maintaining an equilibrium shape.
Note that these are both normalized to $T_{\chi}$ corresponding to the given fBE solution in order to highlight the effect of the non-thermal shape only. 
This can be contrasted with the blue line including the impact of the temperature change as well, where the comparison is made to $\langle \sigma v\rangle_{T}$.

\begin{figure}
    \centering
    \includegraphics[scale=0.8]{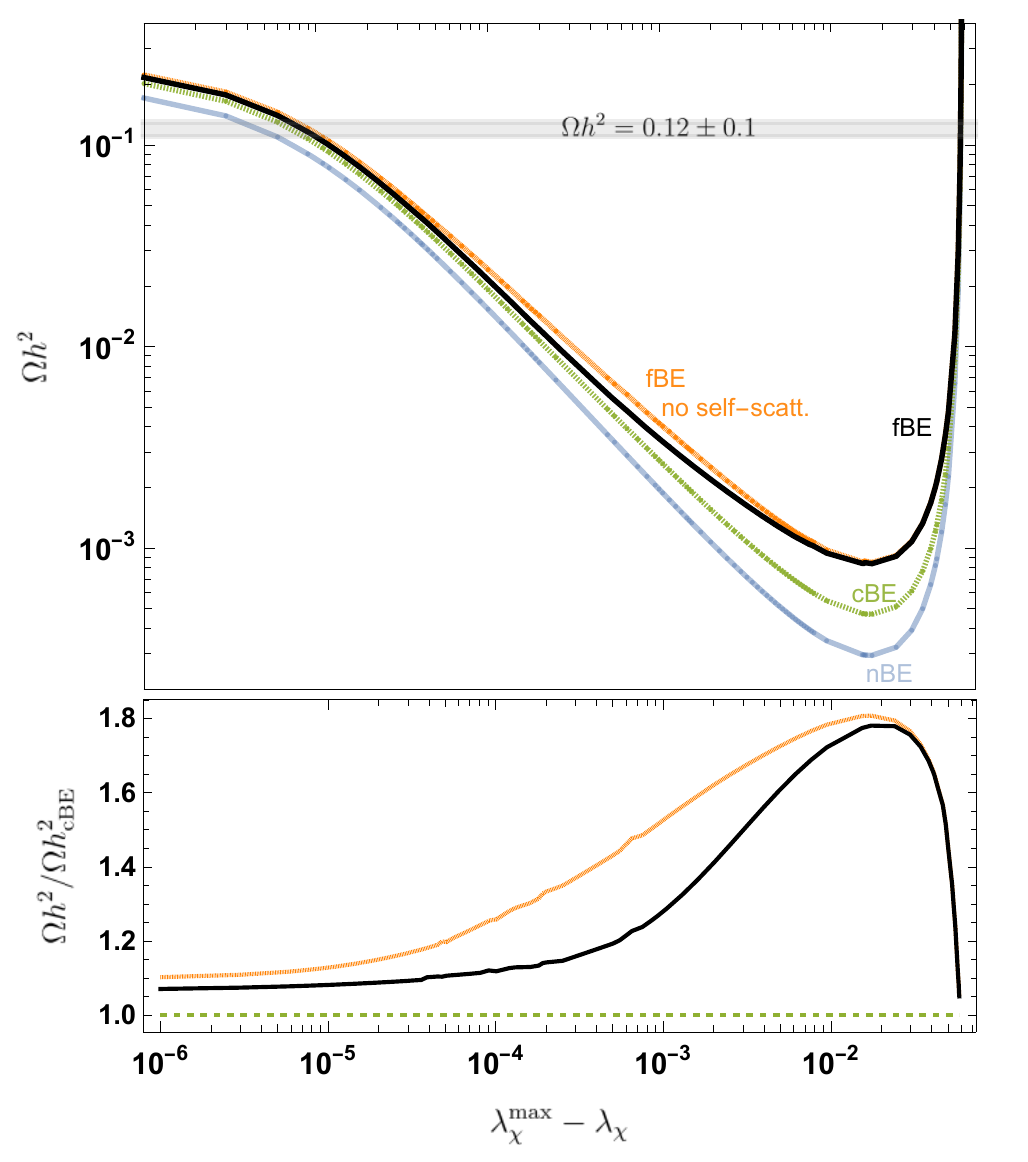}
    \caption{The relic density as a function of the self-coupling for the same benchmark vector resonance model. The blue line gives the result with nBE treatment, green the cBE, orange fBE without self-scatterings, and black of the full calculation. Bottom panel shows the ratio of the last two solutions to the cBE one in order to highlight the size of the effect of self-scatterings.}
    \label{fig:VRES_l}
\end{figure}

Finally, Fig.~\ref{fig:VRES_l} shows how the change of the coupling $\lambda_\chi$ affects the relic abundance for the same benchmark scenario. If this coupling is not fixed by the relic abundance requirement, then fixing $\tilde \gamma$ introduces relation $\lambda_f(\lambda_\chi)$ and one can vary the strength of self-scatterings. Note though, that there is a maximal value, in case of this benchmark point $\lambda_\chi^{\rm max} = 5.855\times 10^{-2}$, above which it is not possible to obtain width $\tilde \gamma = 3\times 10^{-5}$. In order to highlight the most interesting region, in which self-scatterings are as effective as possible, the
$x$-axis of the figure displays the distance from this maximal value $\lambda_\chi^{\rm max}$. Top panel shows the result for the relic density, while the bottom one – the ratios of the result for $\Omega h^2$ obtained with fBE without (orange) and with self-scatterings (black) to the cBE one. For small values of $\lambda_\chi$ (on the right edge of the plot) the fBE result coincides with the one without self-scatterings whatsoever, while for the values approaching the maximal one (on the left edge) it departs towards the result of cBE. 

All in all, these results suggest that for typical values of the self-scattering strength the fBE as currently implemented in DRAKE gives a better approximation of the actual result than the cBE approach. However, this statement is model dependent and when a precise result for the relic abundance is called for, one should in principle fully include self-interactions.

\subsection{Sub-threshold model}
\label{sec:TH}

A second common scenario where the DM annihilation cross-section has a strong velocity dependence is when the DM annihilation process has a threshold at $s>4m_{\rm DM}^2$. Below this threshold the annihilation can be kinematically impossible, sometimes dubbed `forbidden' 
DM~\cite{DAgnolo:2015ujb}, or non-zero, but suppressed. Again for concreteness let us take exactly the same model as the sub-threshold model in Ref.~\cite{Binder:2021bmg}, composed of two scalar particles, where  $\phi_1$ takes the role of the DM, while $\phi_2$ is in thermal contact with the heat-bath fermions $f$. The interaction Lagrangian is given by 
\be
\label{eq:L_threshold}
\mathcal{L} \supset -\frac{g}{4} \phi_1^2 \phi_2^2 -\frac{\lambda}{4!} \phi_1^4 +y_f \phi_2\bar f f\,,
\ee
where compared to the discussion of this model in Ref.~\cite{Binder:2021bmg} we added the self-interaction term, that was not implemented previously. We will assume that the scalars are close in mass and the DM is slightly lighter, i.e., $r\equiv m_2/m_1\gtrsim 1$. In such regime, to the lowest order, the total DM annihilation takes place through the process $\phi_1\phi_1\to\phi_2\phi_2$ and the cross section is given by
\be
\sigma v_{\rm lab} =  \frac{g^2}{32 \pi }  \frac{\sqrt{1-4m_2^2/s}}{s-2 m_1^2}\,,
\label{eq:sv_thr}
\ee
while the momentum transfer rate between the DM and the heat-bath fermions $f$ is strongly suppressed~\cite{Binder:2021bmg}. The amplitude squared of self-scattering in this case is simply a constant and equal to $\lambda^2$.

\begin{figure}
    \centering
    \includegraphics[scale=0.93]{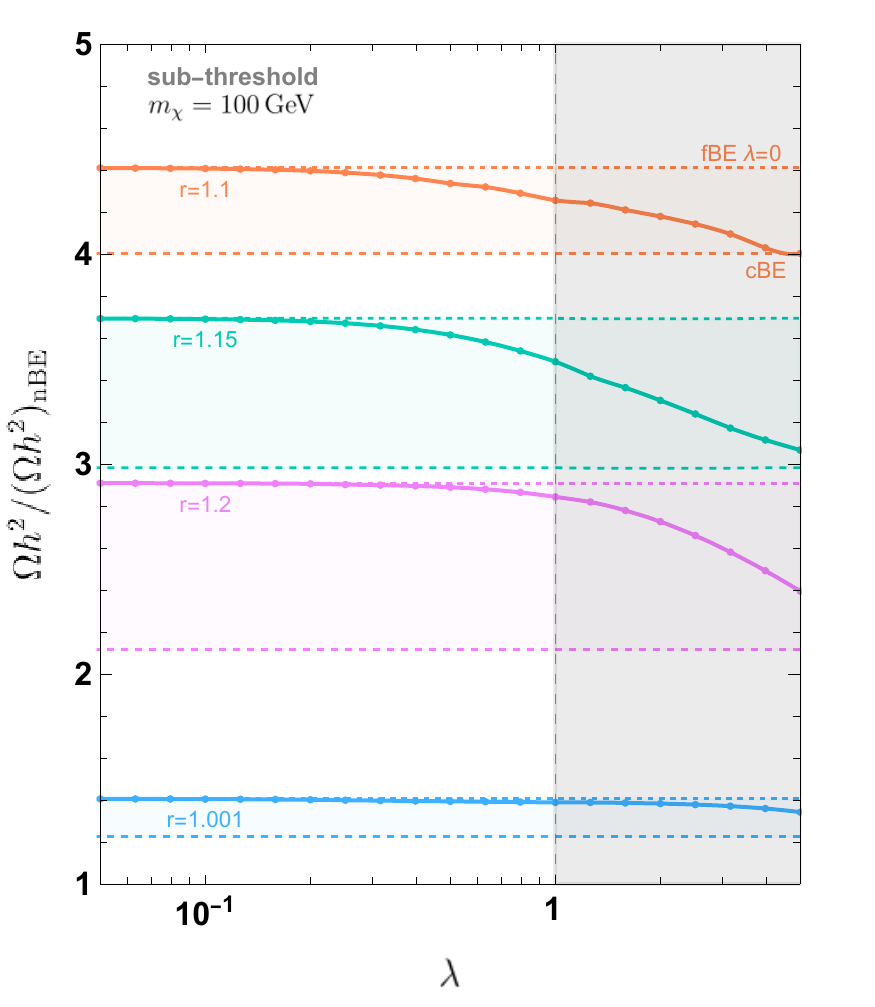}
    \caption{Impact of DM self-scattering on the relic density in the sub-threshold model. For all the chosen values of $r\in\{1.001,1.1,1.15,1.2\}$ the solid line shows the ratio of the full $\Omega h^2$ to the nBE one with the coupling $g$ being fixed by $(\Omega h^2)_{\rm nBE}=0.12$. The dotted lines indicate the corresponding results for cBE and fBE without self-scatterings. The gray shading highlights the region where self-coupling becomes non-perturbative.}
    \label{fig:TH_l}
\end{figure}

In the resonance example we discussed in detail one benchmark model point – here instead let us focus on the relic density as a function of self-scattering strength for a representative set of parameter points. In Fig.~\ref{fig:TH_l} we show the effect on the relic density for four values of $r\in\{1.001,1.1,1.15,1.2\}$ using cBE (dotted), fBE without self interactions (dashed) and full calculation (solid) as a function of the self-coupling $\lambda$. Increasing the self-scattering strength makes the full result go from being the same as fBE $\lambda=0$ to approaching the cBE result, which agrees with the expectations. However, performing the analysis at the quantitative level reveals that, at least for a forbidden-like sub-threshold model at hand, one would require $\lambda > 1$ to actually significantly depart from the fBE $\lambda=0$ result. This confirms the observation from the previous section, that the fBE result from DRAKE is expected to be typically a better estimate for the relic density, than the cBE one. 

Before closing this section a comment on the technical side of the numerical computation is in order. Eq.~\eqref{Cself_def} after discretization
takes the form of a 3 dimensional matrix of entries being 2 dimensional angular integrals, see Eq.~\eqref{eq:disc_Cself}. Therefore, the mitigation of numerical inaccuracies by discretizing on a denser grid is rather costly CPU-wise with $\mathcal{O}(N^3)$ scaling with the grid density for tabulation and also $\mathcal{O}(N^2)$ scaling for generating the collision term matrix at every $x$-step of fBE. Even though typically this matrix is rather well-behaved, this problem becomes especially relevant when the annihilation predominantly relies on the high momentum tail of the distribution. This is the case for the sub-threshold model where the tail is much more prone to numerical error than the bulk of the distribution. All in all, the results presented on the plot in Fig.~\ref{fig:TH_l} required several hours of CPU time per point and still have small, but visible irregularities in the full result at large values of $\lambda$. This attests to the level of numerical accuracy that one can achieve within a manageable CPU cost with our current implementation.\footnote{The numerical code we used is available on request and is planned to be publicly released as an additional optional package to a future DRAKE version.}

\section{Self-thermalization of a non-thermal component}
\label{sec:decay}

The effects of self-scattering are expected to be particularly important for scenarios in which a considerable portion of DM is produced non-thermally on top of the thermal component. Quite commonly particle physics models with the DM candidate(s) contain heavier particles that can decay with a production of one or several DM states. If the lifetime of these heavy particles is sufficiently long, the contribution from the decays does not simply annihilate away and return the distribution to the equilibrium, but can noticeably alter the evolution of DM distribution, its density and other properties \cite{Arcadi:2011ev,Moroi:2013sla,Davoli:2019tpx,Dutra:2021lto,Borah:2021rbx}. This additional non-thermal contribution can also arise, for example, from the bubble collisions following a first-order phase transition \cite{Falkowski:2012fb} or primordial black holes evaporation \cite{Barman:2021ost}. 
If the velocity-averaged cross section of DM annihilation is essentially momentum-independent this injection will have an impact on the rate of annihilation solely by the increase of the density – the resulting relic density will be determined by the interplay of the prolongation of annihilation that depletes the density and the continuous supply of new particles that increases the relic abundance. However, if the annihilation is strongly velocity dependent the effect of the distribution on the annihilation rate is more complicated. If the injected component is rather energetic w.r.t. the thermal one, self-scattering processes will lead to the redistribution of DM particles into the region of the phase space with a larger momentum and hence it can noticeably affect the velocity-averaged cross section.
Moreover, the effects of self-scattering on the energy distribution of DM with a non-thermal component can have consequences that go beyond just the impact on the relic density. For instance, the shape of the relic distribution of light DM particles at later stages of the evolution of the Universe can affect the formation of large scale structure (e.g. \cite{Lin:2000qq,Merle:2015oja,Decant:2021mhj}). However it is worth exploring, the analysis of this phenomenon stays out of the scope of our paper. 

\subsection{Model setup}

To study the self-thermalization of a non-thermal component we take a sterile-neutrino-like model of DM that is coupled to a scalar singlet field $S$, which has been previously studied in the literature (e.g. \cite{Kusenko:2006rh,Petraki:2007gq,Merle:2015oja}), with an additional $U'(1)$ gauge interaction (dark electromagnetism). 
In Ref.~\cite{Ala-Mattinen:2022nuj} a similar model is considered in the context of the impact of non-thermal processes on the DM distribution function. However, it does not include the additional $U'(1)$ making the self-scattering processes absent in their case.
The Lagrangian of the model looks as follows 
\begin{multline}
\mathcal{L} = \mathcal{L}_{\rm SM} + \frac{1}{2}(\partial_{\mu} S)^2 - V(S,H) + yS\Bar{\chi}\chi + m_{\chi}\Bar{\chi}\chi  \\ + \Bar{\chi} i\mathcal{D}_{\mu} \gamma^{\mu} \chi  - \frac{1}{4} F'_{\mu\nu} F'^{\mu\nu} - \frac{\epsilon}{2} F'_{\mu\nu} F^{\mu\nu} + \frac{1}{2} m_A^2 A'_{\mu} A'^{\mu} \, ,
\end{multline}
where $\epsilon$ is the mixing parameter between the photon and the dark photon $A'_{\mu}$, $\mathcal{D_{\mu}} = \partial_{\mu} - ie'A'_{\mu}$ and $V(S,H)$ is the $Z_2$ symmetric part of the scalar potential
\begin{multline}
V(H,S) = -\mu^2_H |H|^2 - \frac{1}{2}m_S^2 S^2 + \lambda_H |H|^4 + \\ + \frac{\lambda_S}{4} S^4 + \frac{\lambda_{HS}}{2} |H|^2 S^2 \, .
\end{multline}
The reason we impose the nearly exact, as explicitly broken only by the interaction of $S$ with DM fermions, $S \rightarrow -S$ symmetry is twofold. For one it motivates the choice of the effective Yukawa coupling $y$ to be small and lead to a long lifetime of $S$ (we will focus on a case where $m_S$ is sufficiently large to allow the decays to a pair of energetic DM particles). But also it allows for a more clear discussion of the process studied in this work without being sidetracked to other, well known effects. In particular, significant explicit or spontaneous breaking of this $Z_2$ symmetry would lead to terms in the Lagrangian that would allow for (loop suppressed, though typically quite efficient) $S$ decay to the SM states. This would not affect directly the discussion of self-thermalization and the size of the DM component coming from $S$ decay could always be adjusted by modifying the $S$ freeze-out. Nevertheless, it could lead to a substantial entropy increase due to the decays of $S$ to radiation and introduce another impact on the relic density of DM that has been already very well studied in the literature (see e.g. \cite{Lazarides:1990xp,Patwardhan:2015kga,Evans:2019jcs}).

In the case that we consider the mass of the dark fermion $m_{\chi}$ is introduced explicitly by the mass term. 
We do not stipulate here the exact mechanism by which the dark photon gets its mass or a UV complete theory of the mixing between the two gauge interactions, however several possible scenarios have been studied in the literature and we refer the reader to a recent review of these models \cite{Fabbrichesi:2020wbt}.

We focus on a region of the model parameters that satisfy the following conditions: \textit{a}) the coupling of the Higgs field to the singlet scalar $S$ is strong enough to keep it in kinetic equilibrium with the SM bath around the DM freeze-out; \textit{b}) $m_A < 2m_{\chi}$, so that the dark photon can only decay to SM states. Then the contribution to the collision term of $\chi(p)$ from the decay of $S(k)$ is given by
\begin{eqnarray}
    C_{\rm dec}  &=&\frac{1}{2g_\chi}\int\!\frac{d^3\tilde p}{(2\pi)^32\tilde E}\int\!\frac{d^3k}{(2\pi)^32\omega} (2\pi)^4\delta^{(4)}(k\!-\!p\!-\!\tilde p)
    \nonumber\\
    &&\MM_{S \leftrightarrow \bar\chi\chi}
f_S(\omega)(1-f_\chi(E))(1-f_\chi(\tilde E)),
\end{eqnarray}
where we have neglected the inverse decay process, as is appropriate for a long-lived $S$. Using the fact that $f_S(\omega)\propto f^{\rm eq}_S(\omega)$ and that $f_\chi(E)\ll 1$ the term above simplifies significantly and is completely independent on the DM distribution function. For the decay amplitude squared 
\begin{equation}
    \MM_{S \leftrightarrow \bar\chi\chi} = \frac{y^2}{2}\left(s-4m_{\chi}^2\right)
\end{equation}
after performing the integrations one arrives at
\begin{eqnarray}
    C_{\rm dec}  =\frac{y^2}{32\pi g_\chi} \frac{n_S}{n_S^{\rm eq}}\frac{T}{pE} \left[e^{-E/T} \kappa(E) \right]_{E_{\rm min}}^{E_{\rm max}}
\end{eqnarray}
with
\begin{eqnarray}
    \kappa(E) &=&4m_{\chi}^2-E^2-2ET-2T^2\, , \\
    E_{\rm max/min}& =& \frac{m_S^2}{2m_{\chi}^2}\left(E \pm p\sqrt{1-\frac{4m_{\chi}^2}{m_S^2}} \right).
\end{eqnarray}
The normalization of this decay term is proportional to the number density of $S$ particles, which can be in or out of equilibrium. It is in turn determined by the chemical decoupling of $S$ from the thermal bath that is governed by its annihilation processes. In our numerical implementation we solve an nBE-type Boltzmann equation for its evolution including the decay process, but neglecting back reaction of $\chi$, which is a very good approximation as long as the decay happens somewhat later than its decoupling. 

The self-scatterings and elastic scatterings are mediated by the dark photon $A'_{\mu}$ 
in the same fashion as in the model considered in Sec.~\ref{sec:VRES}. Depending on the ratio of the masses of DM and the dark photon $m_{\chi}/m_A$ this model can reproduce both of the DM annihilation patterns considered in Section~\ref{sec:ImpactFO}. In the region of masses $m_A/m_{\chi} \lesssim 2$ the DM annihilates to SM electrically charged states via the resonantly-enhanced $s$-channel mixing between the two photon mediators (resonance regime). In the region $m_A/m_{\chi} \gtrsim 1$ annihilation channel to two dark photons is opened
when $s > 4m_A^2$. Far from any resonance, annihilation cross section to dark photons is proportional to $e'^4$, while annihilation to SM states via photon mixing is suppressed by the factor $\epsilon^2\alpha/e'^2$, where $\alpha$ is the fine structure constant. Since the upper limit on the kinetic mixing for the value of the dark photon mass that is relevant for our study is $\epsilon \sim 10^{-3}$ (see e.g. Fig.~3.3 in Ref.~\cite{Fabbrichesi:2020wbt}), the difference of annihilation rates below and above the threshold of $s = 4m_A^2$ is significant, effectively leading to the sub-threshold regime. Despite the smallness of $\epsilon$ the dark photon remains in chemical equilibrium throughout the freeze-out of DM. 

Finally, before discussing the results a comment on the elastic scatterings on bath particles is in order. We follow the implementation in DRAKE, briefly summarized in the Appendix~\ref{append_A}, which was derived in a semi-relativistic Fokker-Planck approximation applicable to the thermal freeze-out. In the presence of an additional relativistic component formally this treatment breaks down and the implemented elastic-scattering term may deviate from the actual one. Nevertheless, as we will see, for the results shown the momentum transfer rate in elastic scatterings, $\gamma(T)$ is subleading compared to the other processes and therefore the current implementation is expected to be sufficient.

\subsection{Results}

Below we show numerical results only for the sub-threshold regime, because the resonance regime for the given model with the constraints on $\epsilon$ requires an extremely sharp resonance to not overproduce the relic density, and so it is phenomenologically less interesting within the model at hand, while retaining similar qualitative behaviour.

\begin{figure}
    \centering
    \includegraphics[scale=0.655]{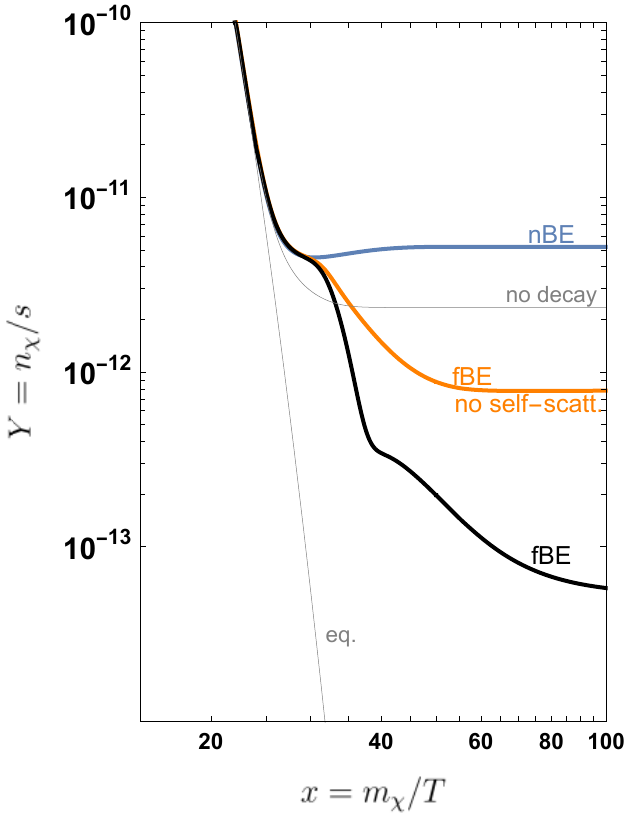}
    \hspace{-0.2cm}
    \includegraphics[scale=0.63]{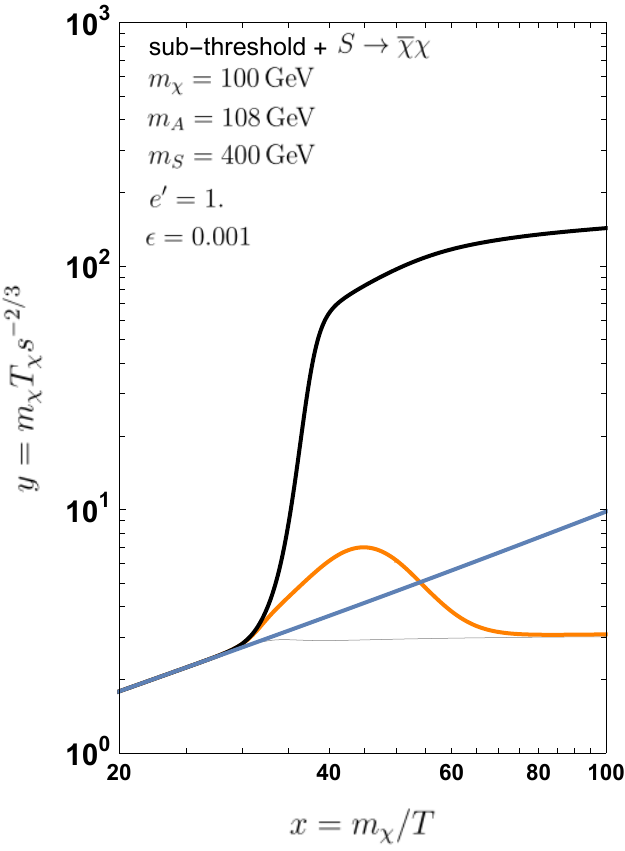}
    \caption{An example evolution of the particle yield $Y$ (left panel) and the temperature parameter $y$ (right panel) for a benchmark sub-threshold+decay model. On both panels the blue line gives the result with nBE treatment, orange fBE without self-scatterings, and black of the full calculation. For comparison gray lines show the equilibrium yield and the one obtained in a model without decay.}
    \label{fig:TH2_Yy}
\end{figure}

\begin{figure}
    \centering
    \includegraphics[scale=0.78]{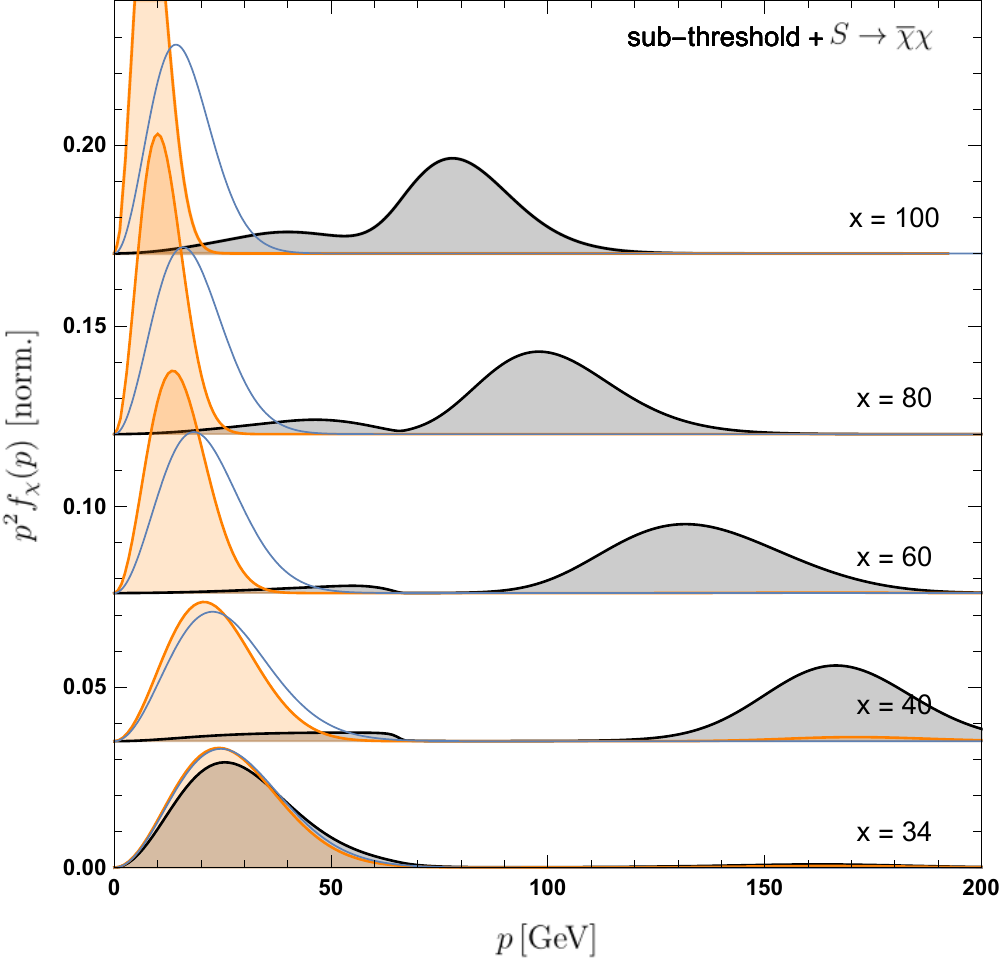}
    \caption{Time snapshots around the freeze-out of the evolution of the normalized momentum distribution for the benchmark sub-threshold+decay model. Black (orange) lines show $f_\chi(p)$ with (without) self-scatterings, while for comparison blue line shows equilibrium distribution at the SM plasma temperature.}
    \label{fig:TH2_f}
\end{figure}

An example of the particle yield $Y$ and the temperature parameter $y$ is shown in Fig.~\ref{fig:TH2_Yy} for a benchmark set of parameters $m_{\chi} = 100 ~\GeV$, $m_A = 108 ~\GeV$, $e' = 1$ and $\epsilon = 0.001$, which is chosen such that the nBE approach with the  decays switched-off (gray curve) reproduces the observed relic density.
All the curves, except for the gray one, display the same inflection point after the density decouples from the equilibrium value. At this point the rate of particle loss due to annihilation and the rate of particle gain due to the decay are comparable. From this point the nBE curve (blue) grows somewhat, but the fBE curves (orange and black) decrease even further. While the behaviour of the first curve is expected as the production of additional DM particles should increase the density, the fact that the account for the actual shape of DM momentum distribution in this case leads to the decrease of the DM density can seem quite surprising. However, it can be easily understood from the velocity-dependent annihilation pattern of the model and the momentum distributions that correspond to the different approaches. The evolution of $f_\chi(p)$ is shown in Fig.~\ref{fig:TH2_f}. In the nBE approach the shape of the distribution is assumed to be unchanged even if the DM particles from $S$ decay are in fact more energetic, hence the rate of annihilation is only slightly affected by the presence of additional DM particles
and the decay gain term dominates over annihilation in the density evolution until it becomes too small to noticeably increase the abundance of DM. In the fBE case with the switched-off self-scatterings (orange) the decays create a small bump in the distribution function with a characteristic momentum that is sufficient to overcome the annihilation threshold and thus the rate of annihilations in the DM gas is significantly boosted. The effective temperature of DM rises w.r.t. the SM plasma temperature (orange line in the right plot in Fig.~\ref{fig:TH2_Yy}), however as the most energetic particles in the bump annihilate away the effective temperature drops below the equilibrium temperature due to the cooling caused by the expansion. 
At this stage the majority of annihilation processes happens below the threshold again, hence the thermally averaged cross section becomes very small and the DM freezes out.

\begin{figure}
    \centering
    \includegraphics[scale=0.885]{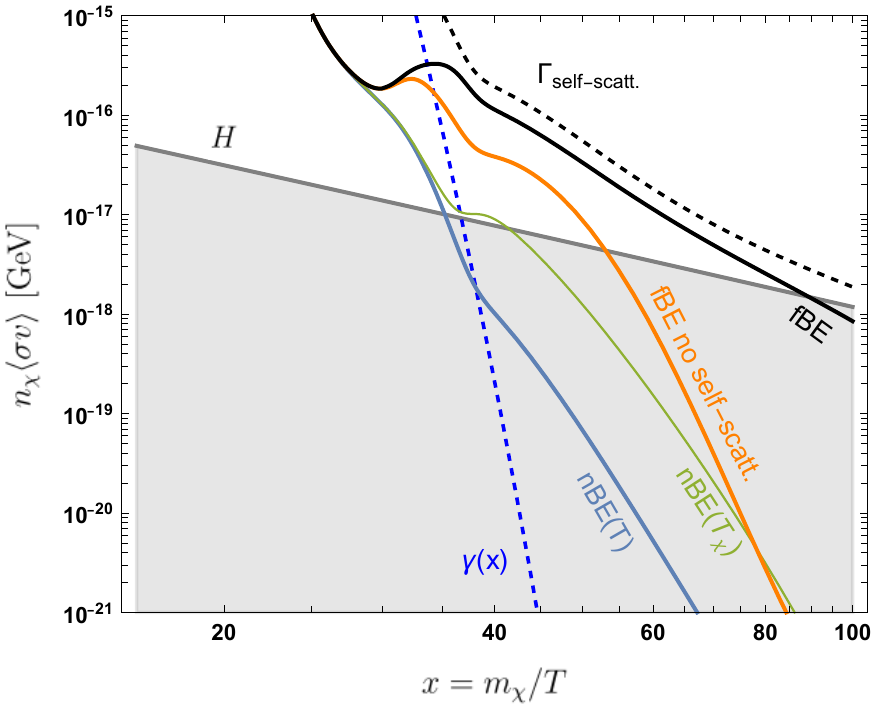}
    \caption{All the relevant rates as compared to the Hubble rate $H$ for the benchmark sub-threshold+decay model: full fBE (black), fBE without self-scatterings (orange), nBE at $T_{\chi}$ and SM plasma $T$ (green and blue, respectively). Dotted blue and black lines show the momentum transfer and self-scattering rates respectively. Note that these are the rates are obtained using the actual solutions for $n_\chi$, not its equilibrium value. The raise of the annihilation rates in the fBE approaches starting at $x\sim 30$ explains why the injection of extra DM particles can lead to a significantly enhanced annihilation and an ultimately decreased relic abundance. The green curve is shown to highlight the fact that it is not the change in temperature, but the shape of the distribution that is the main reason of this enhanced annihilation.
 }
    \label{fig:TH2_sv}
\end{figure}

In the presence of self-scattering the fBE curve (black) displays a steeper drop of the particle yield as the injected DM component heats up the DM gas via elastic collisions such that more DM particles have the energies to overcome the annihilation threshold. In the example that we consider the rate of self-scatterings is large enough compared to the rate of annihilations\footnote{Note, that the intensity of annihilation in the sub-threshold regime and the intensity of self-scattering are controlled by the same coupling.} in the absence of self-scattering (cf. Fig.~\ref{fig:TH2_sv}) that the injected component can effectively transfer the energy to the thermal component before that additional component is annihilated and the energy stored in it is dumped into the SM plasma. This heat significantly increases the temperature of DM and keeps the annihilation rate larger than the Hubble rate for a longer time, so that the relic density is established later and gets a smaller value compared to the case when self-scatterings are switched off. The second inflection point in the abundance curve appears when the supply of DM particles from the decays essentially ceases and little heat is injected in the DM gas. 
In spite of that the rate of annihilation does not decrease as fast as in the absence of self-scattering, because these scatterings promote the clustering of particles in the phase-space region of higher momenta than the characteristic thermal scale and hence prolong the annihilation above the threshold. Note that the enhancement of annihilation that we consider here is only relevant for the early history of DM evolution – long before the effects of DM annihilation can leave an imprint on cosmological observables the DM gas is cooled down by the expansion of the Universe to an extent that annihilation proceeds only below the threshold. 

To summarize, we have considered a practical example of a DM model with a late non-thermal component and demonstrated that the offset between the relic densities predicted by the nBE and fBE treatments can reach up to a few orders of magnitude and that the impact of self-scatterings on the result is crucial as well.  

The above discussion also highlights that there is no good way of formulating a general and simple model independent criterion as to when the DM distribution's departure from equilibrium shape significantly affects the relic density. The final effect comes from an interplay of not only the expansion, annihilation, elastic and self-scattering processes, but also potentially other processes that disrupt the equilibrium, e.g. decays or annihilations of heavier particles into DM. Moreover, comparison of just the rates for these processes is insufficient, as they can be very efficient for some range of the momenta while not for others, as exemplified in the discussed models. A useful rule of thumb of when to expect a possibility of departure from kinetic equilibrium is if the rate of a process that disrupts kinetic equilibrium is larger than the rate of elastic scatterings between the DM and bath particles. As to the importance of careful inclusion of self-scattering processes one can expect it to be necessary when the result obtained with cBE and fBE methods differs to larger extent than the desired accuracy.

Before ending this section let us mention that although we have not studied phenomenological implications of the observed effects, other than the relic abundance, we did check that the presented benchmarks within the example models are feasible. In particular, the present day annihilation cross section is well below the current observational limits and $A'$ decay lifetime is short enough not to spoil the cosmic microwave background anisotropies, nor the Big Bang Nucleosynthesis.

\section{Conclusions}
\label{sec:conclusions}

In this paper we investigated the impact of the DM elastic self-scattering process on the evolution of its momentum distribution function and formation of the relic abundance. 
Building upon the DRAKE framework and code we implemented numerically the self-scattering collision term and applied it to two example models with thermal freeze-out and one model with an additional source of DM particles from decay of a heavier long-lived state. 

We found that in all these cases the effect of self-scattering on the thermalization of the distribution function is large enough to bring visible changes in the effective annihilation rates and therefore final relic abundance of DM. In the freeze-out models our result does follow the expectation of interpolating between the relic abundance obtained using the coupled system of Boltzmann equations for number density and temperature (cBE) and the numerical solution of for $f_\chi(p)$ without including the self-scattering collision term (fBE). This not only validates both of these approaches in their respective limits, but also shows that very large self-scattering rates are needed to recover the cBE result, thus suggesting, though in a model-dependent way, that the fBE is typically a more accurate approach.

In the case of an additional source of DM particles from a decay of a heavier state the disruption of the close-to-equilibrium shape of the thermal component that follows can be very significant, making self-scatterings a crucial ingredient in obtaining the accurate predictions for the relic abundance, temperature and the shape of the DM distribution function. Additionally, we found an intriguing feature that can only be uncovered when studying the evolution at the level of the distribution function: the injection of high energetic DM particles on top of the freeze-out thermal component can lead to a \textit{decrease} of the resulting relic abundance. This effect arises in the scenarios with strong velocity-dependent annihilations and is significantly enhanced due to DM self-interactions. It would be interesting to consider phenomenological implications of such an effect in concrete DM models, which we leave for future work.

Though in this paper we focused on the relic abundance, the provided study also has consequences for the prediction of the matter power-spectrum for light DM candidates that have some non-thermal component. In such situations the precise shape of the distribution function is needed to accurately predict the size of density perturbations and as a consequence can affect, e.g., the warm DM mass limit \cite{Decant:2021mhj}. Our work shows that self-scatterings can play a significant role in such calculations.

\acknowledgments
We would like to thank Tobias Binder for useful discussions. This work is supported in part by the National Science Centre, Poland, research grant No. 2018/31/D/ST2/00813.

\appendix

\section{Numerical implementation of self-scatterings}
\label{append_A}

The collision term for self-scattering from Eq.~\eqref{Cself_def} (taking into account Eq.~\eqref{M2_self}) can be formulated in terms of the discretized momentum components of the distribution function $f_n \equiv f_{\chi}(p_n)$ ($n = 1 \ldots N$) for the numerical implementation on a uniform grid. The integration over momentum $\vec{\tilde{k}}$ can be performed using the 3-dimensional delta-function that imposes the momentum conservation. The residual delta-function can be used for one angular integration, so one is left with two integrals over angles and two integrals over the absolute values of momenta $p$ and $\tilde{k}$, which are approximated by the weighted sums of the corresponding discretized momentum components. Thus, the collision term for self-scattering can be calculated as follows
\begin{multline}
	\label{eq:disc_Cself}
	C_{\rm self} \, [f_i] \approx \frac{(\Delta p)^2}{2 g_{\chi} } \sum_n \sum_m \; F(p_n,p_m) \Big[ f_n \, f_m - f_i \, \tilde{f_j} \Big] \, ,
\end{multline}
where $\Delta p$ is the momentum discretization step-size on the uniform momentum grid, $\tilde{f}_j$ corresponds to the momentum $\tilde{p}_j = \sqrt{p^2_n + p^2_m - p^2_i}$, which is fixed by the momentum conservation. The value of $\tilde{f}_j$ is the result of a linear interpolation between $f_j$ and $f_{j+1}$, which correspond to the nearest momentum nods $p_j \leq \tilde{p}_j \leq p_{j+1}$.
The function $F(p_n,p_m)$ is a double integral over the angles 
\begin{multline}
	\label{Cself_func_F}
	F(p_n,p_m) = \frac{1}{4(2\pi)^4} \int d\cos \theta_2 \int d\cos \theta_3 \\ \times \frac{\MM (s,t)}{\sin\theta_2\sin\theta_3\sqrt{1 - \cos^2\phi}}.
\end{multline}
Here $\theta_3$ is the angle between momenta $\vec{p}_i$ and $\vec{p}_m$ and $\theta_2$ is the angle between momenta $\vec{p}_i$ and $\vec{p}_n$ ($\theta_2$ and $\theta_3$ span in the range from $0$ to $\pi$) and $\phi$ is the angle between the projections of $\vec{p}_n$ and $\vec{p}_m$ onto the plane that is orthogonal to $\vec{p}_i$. The cosine of $\phi$ is fixed by the energy conservation law and can be expressed through the energies and integration angles as follows
\begin{multline}
	\cos\phi = \big(m^2 + E_nE_m - E_nE_i - E_mE_i + \\ 
	+ p_i(p_n\cos\theta_2 + p_m\cos\theta_3) - \\
	- p_np_m\cos\theta_2\cos\theta_3\big)/p_np_m\sin\theta_2\sin\theta_3 \, .
\end{multline}
The actual limits of integration over $\cos\theta_2$ and $\cos\theta_3$ are restrained by the condition
that $|\cos\phi| \leq 1$. In particular, this condition implies that the expression under the square root in Eq.~\ref{Cself_func_F} is not negative. $s$ and $t$ variables can be expressed through the angles and energies as follows
\bea
s & = & 2\left((E_n + E_m)E_i  - p_i(p_n\cos\theta_2 + p_m\cos\theta_3)\right) ,\nonumber \\
t & = & 2m^2 - 2E_iE_n + 2p_ip_n\cos\theta_2 .
\eea

We calculate the integrals in $F(p_n,p_m)$ numerically using a nested adaptive Gauss-Kronrod quadrature. We use the same set of functions $F(p_n,p_m)$ for the backward and forward term similarly to the approach considered in 
Ref.~\cite{Hannestad:1995rs} in the context of general $2 \rightarrow 2$ collision processes, which also requires two numerical integrations over the angles, though it is formulated using different notation. Although the forward term in Eq.~\ref{Cself_def} can be reduced further to an even simpler expression, this is not possible for the backward one. In order to achieve the same level of numerical accuracy and establish a better numerical cancellation between the two terms close to the equilibrium point we treat both terms in the way presented above, which allows to combine them before performing the angular integrals. This increases the stability of the time integration of the Boltzmann equation in the stiff regime in which the distribution function just slightly departs from equilibrium.
A detailed description of the approach to calculate a general $2 \rightarrow 2$ collision terms with a separate treatment of the forward and the backward terms can be found in Ref.~\cite{Ala-Mattinen:2022nuj}.
Also, useful expressions for a specific case $f_1 + f_2 \rightarrow f_3 + f_{\chi}$ can be found in Ref.~\cite{Du:2021jcj}. 

\section{Useful expressions}
\label{append_B}

 In the semi-relativistic regime for DM velocities and small momentum transfer w.r.t. its mass the elastic scattering collision term (Eq.~\ref{Celd_ef}) can be expressed in the Fokker-Planck type approximation \cite{Binder:2016pnr} as
\begin{align}
C_{\rm el} \simeq \frac{E}{2} \gamma(T)
{\Bigg [}
T E\partial_p^2 \!+ \left(2 T  \frac{E}{p} \!+\! p \!+\! T\frac{p}{E}\right) \partial_p + 3
{\Bigg ]}f_{\chi}\,,
\end{align}
where the momentum exchange rate $\gamma(T)$ is given by
\be
  \label{cTdef}
  \gamma(T) =  \frac{1}{48 \pi^3g_\chi m_\chi^3} \int d\omega\,g^\pm
  \partial_\omega\left( k^4 
  \left<\left|\mathcal{M}\right|^2\right>_t\right),
\ee
with
\be  
\left\langle\left|\mathcal{M}\right|^2\right\rangle_t 
\equiv \frac{1}{8k^4}\int_{-4{k}_\mathrm{cm}^2}^0
\!\!\!\! dt(-t)\left|\mathcal{M}\right|^2,
\label{eq:sigmaT}
\ee
and ${k}_\mathrm{cm}^2  \!=\! \left(s-(m_\chi-m_f)^2\right) \left(s-(m_\chi+m_f)^2\right)/(4 s)$ evaluated at ${s=m_\chi^2+2\omega m_\chi+m_f^2}$. 

\bigskip

The full expressions for the functions $\beta_1$ and $\beta_2$ that we use in the expression for the self-scattering amplitude squared of Eq.~\ref{M2self_res} are given below:

\begin{widetext}
\begin{multline}
\beta_1 =  \tilde{\gamma}^2 \left(10 \tilde{s}^2+4 \tilde{s} (4 \tilde{t}-1)+2 \tilde{t} (5
    \tilde{t}-2)+1\right)+4 (\delta +1)^2 \tilde{s}^4+4 (\delta +1) \tilde{s}^3 (-\delta +2 (\delta +1) \tilde{t}-4) + \\
    + \tilde{s}^2 \left(\delta
     (\delta +10) +12 (\delta +1)^2 \tilde{t}^2-24 (\delta +1) \tilde{t}+19\right)+\tilde{s} \left(-\delta +8 (\delta +1)^2 \tilde{t}^3-24
    (\delta +1) \tilde{t}^2-(\delta -3) (\delta +5) \tilde{t}-5\right) + \\
    + \tilde{t} \left(-\delta + \tilde{t} \left(\delta  (\delta +10)+ +4
    (\delta +1)^2 \tilde{t}^2-4 (\delta +4) (\delta +1) \tilde{t}+19\right)-5\right)+1 \, ;
\end{multline}
    
\begin{multline}
\beta_2 = -3 \delta +g^2 \left(10 \tilde{s}^2+4 \tilde{s} (\tilde{t}-4)+4 (\tilde{t}-1)
    \tilde{t}+7\right)+4 (\delta +1)^2 \tilde{s}^4+4 (\delta +1) \tilde{s}^3 (2 (\delta +1) \tilde{t}-3 \delta ) + \\
    +\tilde{s}^2 \left(\delta (13 \delta -2)+12 (\delta +1)^2 \tilde{t}^2-12 \delta  (\delta +1) \tilde{t}-5\right)+\tilde{s} (-6 \delta ^2+9 \delta +8 (\delta+1)^2 \tilde{t}^3-12 \delta  (\delta +1) \tilde{t}^2 + \\
    + (\delta -1) (3 \delta +1) \tilde{t}-1)+\delta ^2 ((\tilde{t}-1) \tilde{t} (4(\tilde{t}-1) \tilde{t}-1)+1)+2 \delta  (1-2 \tilde{t})^2 (\tilde{t}-1) \tilde{t}+(\tilde{t}-1) \tilde{t} (4 (\tilde{t}-1)\tilde{t}+7)+3 \, .
\end{multline}
\end{widetext}

\bibliography{biblio}

\end{document}